\documentclass[
10pt,aps,
notitlepage, 
bibnotes,
prb,
twocolumn,
showpacs,preprintnumbers,superscriptaddress,
amsmath,amssymb,floatfix]{revtex4-2}

\usepackage{graphicx}
\usepackage{dsfont}
\usepackage{dcolumn}
\usepackage{bm}
\usepackage{color}
\usepackage{xcolor}
\usepackage{url}
\usepackage{tabularx}
\usepackage{array}
\usepackage{booktabs}
\usepackage{comment}
\usepackage{hyperref}
\usepackage{footnote}
\usepackage{lipsum}
\usepackage[normalem]{ulem}
\usepackage{booktabs}
\usepackage{braket}

\DeclareSymbolFont{md}{OMX}{mdput}{m}{n} 
\DeclareMathSymbol{\intop}{\mathop}{md}{90}

\definecolor{colorA}{rgb}{0, 0, 1}

\makeatletter
  \def\my@tag@font{\normalsize}
  \def\maketag@@@#1{\hbox{\m@th\normalfont\my@tag@font#1}}
  \let\amsmath@eqref\eqref
  \renewcommand\eqref[1]{{\let\my@tag@font\relax\amsmath@eqref{#1}}}
\makeatother

\begin{document}

\def\afflux{Department of Physics and Materials Science, University of Luxembourg, L-1511 Luxembourg, Luxembourg}

\title{Quantum and classical magnetic Bloch points}

\author{Vladyslav M. Kuchkin}
\email{vladyslav.kuchkin@uni.lu}
\affiliation{\afflux}

\author{Andreas Haller}
\affiliation{\afflux}

\author{\v{S}tefan Li\v{s}\v{c}\'{a}k}
\affiliation{\afflux}

\author{Michael P. Adams}
\affiliation{\afflux}

\author{Venus Rai}
\affiliation{\afflux}

\author{Evelyn P. Sinaga}
\affiliation{\afflux}

\author{Andreas Michels}
\affiliation{\afflux}

\author{Thomas L. Schmidt}
\affiliation{\afflux}

\date{\today}

\begin{abstract}
A Bloch point represents a three-dimensional hedgehog singularity of a magnetic vector field in which the magnetization vanishes. 
However, standard micromagnetic theory, developed for magnetic moments of fixed lengths, lacks full applicability in studying such singularities.
To address this gap, we study a Bloch point in a quantum Heisenberg model for the case of spin-1/2 particles.
Performing an exact diagonalization of the Hamiltonian as well as using density matrix renormalization group techniques, we obtain the ground state, which can be used to recover the corresponding magnetization profile.
Our findings demonstrate a variation of the spin length in the quantum model, leading smoothly to zero magnetization at the Bloch point. 
Our results indicate the necessity of generalizing the classical micromagnetic model by adding the third degree of freedom of the spins: the ability to change its length.
To this end, we introduce the micromagnetic $\mathbb{S}_{3}$-model, which enables the description of magnets with and without Bloch point singularities.
\end{abstract}

\maketitle 

\textit{Introduction.} Bloch points (BPs) are micromagnetic singularities which were introduced by Feldtkeller~\cite{Feldtkeller_1965} and D\"oring~\cite{Doring_1968} already in the 1960s.
For a unit magnetization vector field $\bm{m}(\bm{r})$, such a Bloch point represents a statically stable hedgehog solution, $\bm{m}=\bm{r}/r$, with finite energy existing in the classical Heisenberg model. 
Since the magnetization of the BP does not approach a constant ferromagnetic background (e.g., $(0,0,1)$ at $r\rightarrow\infty$), it represents strictly speaking a defect in the magnetic microstructure. Up to date, a variety of spin textures hosting BPs were theoretically predicted and experimentally found. 
The most prominent examples are hard magnetic bubbles with Bloch lines~\cite{Malozemoff_79,Fruchart}, chiral bobbers~\cite{Rybakov_2015}, lattices of bobbers~\cite{Ran} as well as globules~\cite{globule}.
Recently, theoretical studies demonstrated a chain of BPs stabilized by screw dislocations~\cite{Azhar}, which might play a crucial role in the formation of hopfion rings~\cite{hopfion,Nagaosa}.
These spin textures are intensively studied for information storage and as information carriers in future electronic devices~\cite{Beg,Fischer,Guslienko,Aliev}.

At the same time, the dynamics of magnetization textures hosting BPs cannot be correctly described by a straightforward micromagnetic model because the singularity of the effective magnetic field at the center of the BP gives rise to a divergent Heisenberg exchange interaction, $\bm{b}_\mathrm{eff}\sim \nabla^2\bm{m}$.
As a consequence, BP dynamics is mainly studied in atomistic spin models, where strong interactions of the BP with the lattice have been reported~\cite{Kwon,Gong}.
There have been attempts to combine atomistic and micromagnetic approaches by considering multiscale grids~\cite{Hertel} but this is a complex method that requires treating each BP separately.
A more straightforward approach would be to generalize the micromagnetic description of such singularities by allowing magnetization vector fields of variable length.
In previous studies~\cite{Verga,Chubykalo}, such a variation of the magnetization was justified by the presence of thermal fluctuations, which play an important role close to the Curie temperature.
However, the experimentally observed textures hosting BPs do not necessarily require high temperatures, so it seems that the temperature should not play a crucial role. 
In our study, we stay within the framework of a micromagnetic theory, which treats all magnets below the Curie temperature as athermal~\cite{Landau, Hubert}.
\begin{figure}[tb!]
\centering
\includegraphics[width=8cm]{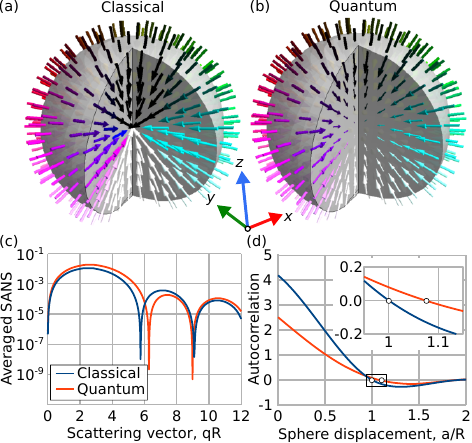}
\caption{~\small Classical and quantum Bloch points.
Panels (a) and (b) show the magnetization vector fields for the classical ($\bm{m}=-\bm{r}/r$) and the quantum ($\bm{n}=-\bm{r})$ hedgehog solutions, respectively, for $r\in[0,1]$.
The same color scheme for the spin visualization is used throughout the paper \cite{Magnoom,Savchenko}.
Panels (c) and (d) display the analytical dependencies of the magnetic neutron scattering cross sections~\eqref{averaged_sans} and their autocorrelation functions obtained for these vector fields (see \cite{Supplement}).
} 
\label{Fig0}
\end{figure}
\begin{figure*}
\centering
\includegraphics[width=18cm]{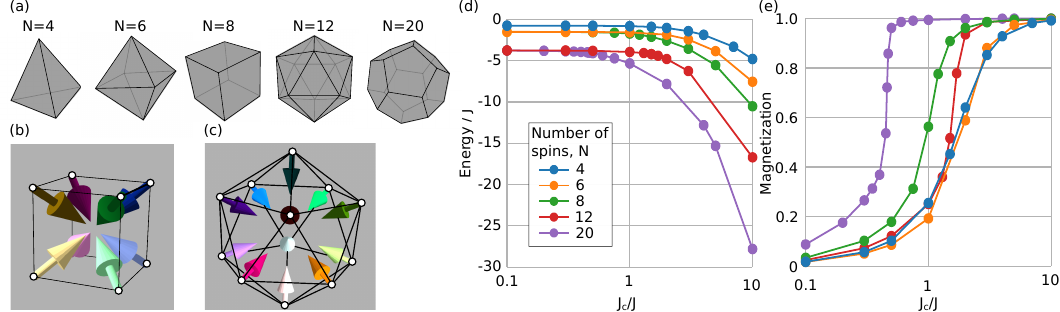}
\caption{~\small Bloch point results for the single-shell case.
Panel (a) shows the Platonic solids corresponding to equidistant points on the sphere.
Panels (b) and (c) display the BP texture for the case of a cube ($N=8$) and an icosahedron ($N=12$), respectively.
Panels (d) and (e) show the energy (in units of $J$) and the magnetization length $|\bm{n}|$ for different coupling parameters $J_\mathrm{c}/J$.} 
\label{Fig1}
\end{figure*}

As we show in this Letter, the problem of a singular effective field near the center of the BP can be resolved if one allows the magnetization to change its length by introducing a new order parameter $\bm{n}$ with $|\bm{n}|\in[0,1]$.
Considering a more general quantum-mechanical Heisenberg model underlying the classical micromagnetic model allows us to demonstrate explicitly such a variation of the spin length.
In this model, the order parameter is the quantum mechanical wave function $\ket{\psi}$, which can be used to calculate the magnetization expectation values $\bm{n}=\braket{\psi|\bm{S}|\psi}/(s\hbar)$ with $|\bm{n}|\in[0,1]$. In the following, we will assume $s=1/2$, but a generalization to other spin values is straightforward.

To distinguish such a magnetization profile from the one obtained by Feldtkeller and D\"oring~\cite{Feldtkeller_1965,Doring_1968}, we refer to these BPs as quantum and classical ones, respectively, and describe them by the order parameters $\bm{n}$ and $\bm{m}$.
We illustrate this in Fig.~\ref{Fig0}(a,b) for the hedgehog BP.
Asymptotically, for stabilization, both BPs must exhibit a $-\bm{r}/r$ dependency for $r\rightarrow\infty$. 
As we show below, in the quantum case, this can be achieved by adding a Zeeman term near the boundary of the BP.
It is worth highlighting that the common property shared by the classical and quantum BPs is their spherical symmetry.
This can be written as $\bm{n}(r)/|\bm{n}(r)|=\bm{m}(r)=-\bm{r}/r$ for $r>0$.
The main difference between the quantum mechanical and classical magnetization textures occurs near the BP core, where one has $\bm{n}(0)=0$, while $\bm{m}(0)$ is undefined.
We shall see below that for magnetic textures without singularities, the order parameters $\bm{m}$ and $\bm{n}$ coincide.

Based on our results for the BP in a three-dimensional quantum Heisenberg model, we propose a generalized micromagnetic model which makes it possible to describe magnetic textures both with and without singularities. 
Since the lattice structure of the quantum-mechanical model tends to break the spherical symmetry of the BP, we will consider a special lattice geometry, based on shells, for solving the quantum model. 
The micromagnetic model obtained in this way is free of this effect and can be applied to different magnetic systems characterized by isotropic exchange interactions. 
As we also show in Fig.~\ref{Fig0}(c,d), experimentally measurable quantities, such as the magnetic neutron scattering cross section and its autocorrelation function, show a clear difference between classical and quantum BPs. 
The details of our calculations are presented below and in the Supplement Material~\cite{Supplement}.

\textit{Model.} We start from the quantum Heisenberg model for a single spherical shell of $N$ interacting spin-$1/2$ particles. Each spin is represented as $\bm{S}_{i} = \hbar\bm\sigma_i/2$ using a vector of Pauli matrices $\bm\sigma_i$, and the Hamiltonian reads,
\begin{eqnarray}
 & \mathcal{H}_{1}=-J{\displaystyle \sum_{\left\langle i,j\right\rangle }\boldsymbol{S}_{i}\cdot\boldsymbol{S}_{j}-J_{c}}{\displaystyle \sum_{i=1}^{N}}\boldsymbol{S}_{i}\cdot\boldsymbol{h}_{i}.\label{Hamiltonian}
\end{eqnarray}
The first term describes an intra-shell ferromagnetic nearest-neighbor exchange coupling with a strength $J > 0$. In order to stabilize spherically-symmetric BPs, we are limited to $N\in \{ 4,6,8,12, 20\} $ for the number of sites per shell, corresponding to the Platonic solids [see Fig.~\ref{Fig1}(a)]. 
Using any other number of spins in a single shell will necessarily break the spherical symmetry and lead to a misorientation of the spins relative to the spherically-symmetric hedgehog field.
The environment of the shell is described by an effective Zeeman term with spherically symmetric magnetic field $\bm{h}_{i}=-\bm{r}_{i}/r_{i}$, where the $\bm{r}_{i}$ refer to the spin positions and $J_c > 0$ is an effective coupling constant. 
\begin{figure*}
\centering
\includegraphics[width=18cm]{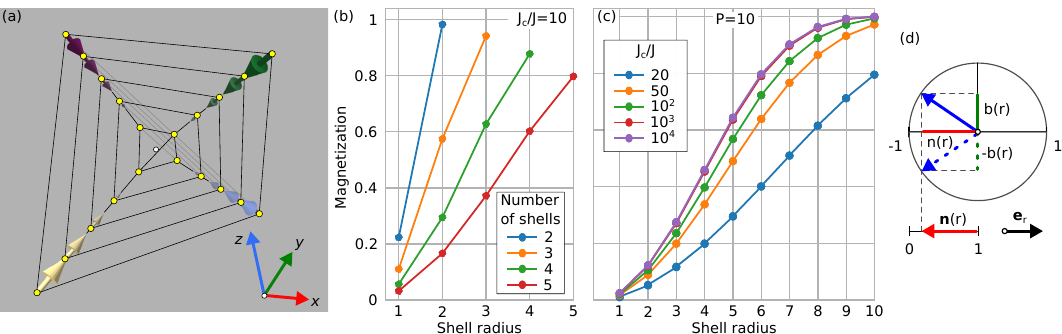}
\caption{~\small BP results for the multi-shell case.
(a) shows the quantum BP spin texture obtained for the tetrahedral case with five shells. Bonds between interacting spins are shown. 
(b) and (c) Dependency of the magnetization length on the shell radius.
(b) corresponds to $P=2,3,4,5$ shells [Eq.~\eqref{Hamiltonian_multi}] with $J_\mathrm{c}/J=10$ (obtained using the ED method). 
(c) corresponds to $P=10$ and $J_\mathrm{s}/J=20,50,10^{2},10^{3},10^{4}$ (obtained using the DMRG method).
(d) demonstrates the mapping of the vector $\bm{n}$ with $n\in[0,1]$ onto the half circle with $n^{2}+b^{2}=1$.
} 
\label{Fig2}
\end{figure*}

In the case of multiple shells, the Hamiltonian~\eqref{Hamiltonian} is generalized to
\begin{align}
    \mathcal{H}_\mathrm{P}=
    -J\!\sum_{\left\langle i,j\right\rangle ,p} \boldsymbol{S}_{i}^{p}\cdot\boldsymbol{S}_{j}^{p} 
    -J^{\prime} \sum_{i,\left\langle p,q\right\rangle} r_{p}^{2}\boldsymbol{S}_{i}^{p}\cdot\boldsymbol{S}_{i}^{q}-J_{c} \sum_{i}\boldsymbol{S}_{i}^{P}\cdot\boldsymbol{h}_{i},\label{Hamiltonian_multi}
\end{align}
where $p \in \{1, \ldots, P\}$ sums over all $P$ shells, $J^{\prime} > 0$ is the strength of the inter-shell interaction, $r_p=p$ is the (dimensionless) radius of the $p$th shell, and the superscript of the spin operators $\boldsymbol{S}_{i}^{p}$ denotes the shell number. 
The last term in Eq.~\eqref{Hamiltonian_multi} involves only spins from the outermost ($p=P$) shell, which are coupled to the external magnetic field to satisfy the boundary condition mentioned above. 
The total number of spins is $NP$ in this case. For the case of isotropic exchange, we set $J^{\prime}=J$ (see below).
More details on the derivation of the multi-shell Hamiltonian~\eqref{Hamiltonian_multi} are provided in the Supplemental Material~\cite{Supplement}.

\textit{Single shell.} The ground state (GS) of the Hamiltonian~\eqref{Hamiltonian} can be obtained by performing an exact diagonalization (ED) for a sufficiently small number of lattice spins.
The corresponding magnetic textures for the cube ($N=8$) and the icosahedron ($N=12$) cases are shown in Fig.~\ref{Fig1}(b,c).
These examples are general, and we observe a similar ordering behavior of magnetic spins for all other cases depicted in Fig.~\ref{Fig1}(a).
At the same time, the dependencies of the energy [Fig.~\ref{Fig1}(d)] and of the spin length [Fig.~\ref{Fig1}(e)] on the ratio $J_\mathrm{c}/J$ are different due to the different number of spins and bonds in each case.
In particular, for $J_\mathrm{c}=0$ the energy depends on the number $N_\mathrm{b}$ of interacting pairs according to $-J N_\mathrm{b}/4$. 
In the strong-coupling limit ($J_\mathrm{c}\gg J$), the energy is of the order of $-J_\mathrm{c}N$.

Moreover, we have analytically studied the simplest case of $N=4$ (tetrahedron), while the other cases shown in Fig.~\ref{Fig1} are physically similar and because of symmetry we do not expect any qualitative difference.
We considered the limiting case of $J_\mathrm{c}/J\rightarrow0$ and calculated the energies of the GS and of the first four excited states in the Supplemental Material~\cite{Supplement}.
From this and from the obtained numerical results in Fig.~\ref{Fig1}, we can deduce the uniqueness of the GS for all $J_\mathrm{c}/J > 0$.
Further analytic investigations can be done with the ansatz for the GS, which we discuss in the Supplemental Material~\cite{Supplement}.
% %

\textit{Multiple shells.} From the discussion of a single shell, we can deduce that the qualitative physics remains the same for each $N$ corresponding to a Platonic solid. Thus, we can use the simplest case $N=4$ to approach the thermodynamic limit $P \rightarrow\infty$ in the multi-shell Hamiltonian~\eqref{Hamiltonian_multi} and to describe the quantum BP in the bulk limit. To solve Eq.~\eqref{Hamiltonian_multi} numerically, we rely on ED and the density matrix renormalization group (DMRG)~\cite{White, Haller} methods.
Both yield identical results for $P\in\{2,3,4,5\}$ while DMRG remains computationally feasible even for $P>5$.

In both cases, the corresponding magnetic texture is the spherically-symmetric quantum BP with a spin length that decreases towards the origin, as highlighted in 
Fig.~\ref{Fig2}(b) for $J_\mathrm{c}/J=10$. 
This example is generic, and the profile $n(\bm r) = |\bm n(\bm r)|$ is always an increasing function of $r = |\bm r|$ with $0 \leq n(r) \leq 1$.

For a larger system with ten shells, we obtain the ground state using DMRG [see Fig.~\ref{Fig2}(c)].
We have examined different values of $J_\mathrm{c}/J$ and we find that in the limit $J_\mathrm{c}/J\!\gg\!1$, the function $n(r)$ approaches a certain limiting curve -- the profile of the quantum BP. 
Guided by these results, we provide next a micromagnetic model that can be used to describe such BPs in systems of arbitrary geometry.  

\begin{figure*}
\centering
\includegraphics[width=18cm]{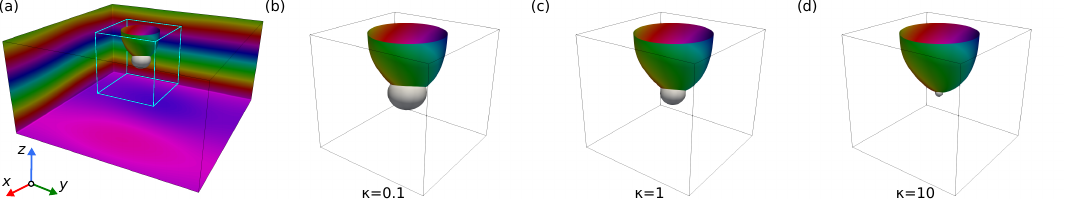}
\caption{~\small Case of a chiral bobber.
Subfigure~(a) shows the relaxed chiral bobber stabilized at a magnetic field of $0.75\mathcal{D}^{2}/2M_\mathrm{s}\mathcal{A}\bm{e}_\mathrm{z}$; periodic and open boundary conditions were, respectively, assumed in the $xy$~plane and along the $z$~direction. The size of the light blue box is $4\pi\mathcal{A}/\mathcal{D}\sim43$ lattice sites, and the shown surface of the bobber is given by $n_\mathrm{z}=0$.
The white ball corresponds to a magnetization length of $n=0.95$, meaning that inside (outside) the ball one has $0\leq n<0.95$ ($0.95<n\leq1$).
(b)$-$(d) show the bobber for different values of the parameter $\kappa$ given in units of $\mathcal{D}^{2}/2M_\mathrm{s}\mathcal{A}$.
} 
\label{Fig4}
\end{figure*}

\textit{Micromagnetic $\mathbb{S}_{3}$-model.} As we have shown above, the quantum BP is characterized by a spherically symmetric magnetization profile with varying length $0 \leq n(r) \leq 1$. 
This inequality constraint can be written as an equality constraint by extending the order parameter to four dimensions by introducing a vector $\bm b$ with the properties
\begin{equation}
    n(r)^{2}+b(r)^{2}=1,\,\,\bm{n}(r)\cdot\bm{b}(r)=0,\label{constraint}
\end{equation}
where $\bm{b}=b(r)\bm{e}_{b}$ and the unit vector $\bm{e}_{b}$ is orthogonal to the spherical basis vectors $\bm{e}_{r}, \bm{e}_{\theta}, \bm{e}_{\phi}$.
As shown in Fig.~\ref{Fig2}(d), both projections $b(r)$ and $-b(r)$ correspond to the same magnetization length $n(r)$.
This is due to the domain of $\bm{n}$ being a ball $\mathbb{B}_{3}$, while the constraint \eqref{constraint} defines a sphere $\mathbb{S}_{3}$.
The strict mapping would be from $\mathbb{B}_{3}$ onto the half-sphere with $b(r)\geq0$. However, one can allow $\bm n$ and $\bm b$ to cover the entire $\mathbb{S}_{3}$ domain and obtain a unique physical solution for the magnetization $\bm{n}$ if the generalized Hamiltonian satisfies $\mathcal{E}(\bm{n},\bm{b})=\mathcal{E}(\bm{n},-\bm{b})$.
Taking that into account, the micromagnetic model we propose can be written in the following form,
\begin{eqnarray}
\mathcal{E}=\mathcal{A}\int\left[\left(\nabla\bm{n}\right)^{2}+\left(\nabla\bm{b}\right)^{2}\right]\mathrm{d}V+\kappa\int b^{2}\mathrm{d}V,\label{model4D}
\end{eqnarray}
where $\mathcal{A}$ is the exchange-stiffness constant, and the last term with the coupling constant $\kappa>0$ is introduced to make it possible to connect to the classical micromagnetic limit by using $\kappa\rightarrow\infty$, which leads to $b^{2}=0$ or equivalently $n^{2}=1$.
Following Ref.~\cite{Doring_1968}, we can integrate the quantum BP energy functional over a ball of radius $R$ and get a result of the form $\mathcal{E}_\mathrm{BP}=8\pi\mathcal{A}R+\Omega$. Here, $\Omega$ represents a negative contribution to the energy arising from the possibility of the spins to change their lengths. Parametrizing the magnetization by an angle $\psi(r)\in[0,\pi/2]$, as $n(r)=\cos\psi$, and by $b(r)=\sin(\psi)$, one has 
\begin{align}
  \Omega=4\pi\mathcal{A}\int_{0}^{R}\left[\left(\psi^{\prime}\right)^{2}+\left(\!\dfrac{\kappa}{\mathcal{A}}\!-\!\dfrac{2}{r^{2}}\right)\sin^{2}\psi\right]r^{2}\mathrm{d}r.  
\end{align}
From the Euler-Lagrange equations one finds the asymptotic limits of $\psi(r)$ as $\psi(r\ll 1) = \pi/2-c_{1}^{2} r/r^{*}$ and $\psi(r\gg1) = (c_{2}^{2} r/r^{*})\exp\left(-r/r^{*}\right)$ where $r^{*}=\sqrt{\mathcal{A}/\kappa}$. 
In \cite{Supplement}, we show that an ansatz of the form $\psi_\mathrm{a}(r)=\arccos[\tanh \left(c r/r^{*}\right)]$ with the only fitting parameter $c$ approximates very well the true quantum BP profile.

The profile of a quantum BP is described by an exponentially decaying function, $|\bm{n}|\propto1-\exp(-r/\sqrt{\mathcal{A}/\kappa})/r$ when $r\gg1$. This implies that the ratio $\sqrt{\mathcal{A}/\kappa}$ can be used to estimate the size of the quantum BP. Moreover, such exponential localization means that model \eqref{model4D} indeed coincides with the standard micromagnetic approach far from the BP singularity and modifies it only near the BP core. 

It is useful to compare the effective field $\bm{b}_\mathrm{eff}$ for classical and quantum BPs.
As we show in the Supplemental Material~\cite{Supplement}, one finds $|\bm{b}_\mathrm{eff}|\sim1/r^{2}$ in the classical case, whereas the quantum case leads to $|\bm{b}_\mathrm{eff}|\sim r$ for $r\rightarrow0$.
Thus, the continuous behavior of the effective field near the quantum BP singularity in principle makes it possible to derive dynamical equations based on the same assumptions as in Landau and Lifshitz's original paper~\cite{Landau}.

\textit{Chiral bobber with a quantum BP.} The simplest experimentally observed spin texture that hosts a BP is a chiral bobber.
This state represents a skyrmion tube protruding into a ferromagnetic medium up to a certain depth and ending with a BP.
In addition to the terms in Eq.~\eqref{model4D}, the Hamiltonian allowing the stabilization of such a state contains a Dzyaloshinskii-Moriya interaction term of strength $\mathcal{D}$~\cite{Dzyaloshinskii,Moriya} (DMI) and an externally applied field.
Choosing the strength of this field within the stability range of a chiral bobber, we study the role of the $\kappa$~parameter in Fig.~\ref{Fig4}.
In agreement with the above results on an isolated BP, this parameter defines the size of the quantum BP (see the white ball).
As $\kappa$ increases, the quantum BP shrinks and transforms into the classical BP solution.
The vector field in the vicinity of the BP core coincides with that shown in Fig.~\ref{Fig0}(a) up to a rotation about the $z$~axis by $\pi/2$.

Thus, the generalized model~\eqref{model4D} can be supplemented by other terms, such as DMI and magnetic fields, relevant to the particular physical case. 
Moreover, dipolar interactions can be straightforwardly included because they do not require the conservation of the spin length.

The main advantage of the $\mathbb{S}_{3}$~model as compared to previously considered models that allow for a spin length variation is that it coincides with the standard micromagnetic approach for continuous magnetic vector fields and corrects it only in the vicinity of BP singularities, so that the magnetization always remains constrained by $|\bm{n}|\leq 1$. 
Furthermore, mathematically the suggested model is based on the same order parameter used in the Skyrme baryonic model~\cite{Skyrm}. Thus, the quantum BP is analogous to a three-dimensional skyrmion in magnetism. 
One can show that the quantum BP discussed here is characterized by a baryonic charge of $1$, and the question of the existence of higher-order BPs will remain a subject for further work.

\textit{Experimental signatures of a quantum BP.} In the simplest case of a spherically symmetric BP, the magnetic neutron scattering cross section can be obtained analytically~\cite{Supplement}. Here, we provide the results for the cross sections of the classical ($I_{1}$) and the quantum ($I_{2}$) BPs in a ball of radius $R$,
\begin{eqnarray}
 && I_{1} \simeq (qR\sin qR + 2\cos qR - 2)^{2}/(qR)^{6} \nonumber\\
 && I_{2} \simeq \left(3qR\cos qR + \left((qR)^{2}-3\right)\sin qR\right)^{2}/(qR)^{8},\label{averaged_sans}
\end{eqnarray}
which are plotted in Fig.~\ref{Fig0}(c).
These expressions and their autocorrelation functions [Fig.~\ref{Fig0}(d)] can be used straightforwardly in analyzing experimental data to distinguish the classical BP from its quantum mechanical counterpart.

\textit{Conclusion.} We have studied the Bloch point (BP) singularity in the quantum Heisenberg model for spin-$1/2$ particles.
The obtained magnetization vector field is characterized by a vanishing spin moment at the BP core.
We suggested a generalized micromagnetic model for this quantum BP based on an $\mathbb{S}_{3}$ order parameter and applied this model to a chiral bobber.
To distinguish experimentally between a classical and a quantum BP, we have provided expressions for the magnetic neutron scattering cross sections and the corresponding autocorrelation functions.

\textit{Acknowledgments.} We acknowledge financial support from the National Research Fund Luxembourg under Grant~C22/MS/17415246/DeQuSky. VMK is grateful to Nikolai Kiselev for fruitful discussions.

\onecolumngrid

\section*{Supplemental Material for ``Quantum and classical magnetic Bloch points''}

\section{Weak coupling limit for $N=4$}
In the tetrahedral case for a single shell, the Hamiltonian can be written as
\begin{align}
  \mathcal{H}_{1}&=-J\left[\boldsymbol{S}_{1}\cdot\boldsymbol{S}_{2}+\boldsymbol{S}_{1}\cdot\boldsymbol{S}_{3}+\boldsymbol{S}_{1}\cdot\boldsymbol{S}_{4}+\boldsymbol{S}_{2}\cdot\boldsymbol{S}_{3}+\boldsymbol{S}_{2}\cdot\boldsymbol{S}_{4}+\boldsymbol{S}_{3}\cdot\boldsymbol{S}_{4}\right]\label{Ham_tetr}-J_{c}\left[\boldsymbol{S}_{1}\cdot\boldsymbol{h}_{1}+\boldsymbol{S}_{2}\cdot\boldsymbol{h}_{2}+\boldsymbol{S}_{3}\cdot\boldsymbol{h}_{3}+\boldsymbol{S}_{4}\cdot\boldsymbol{h}_{4}\right],\\
 \boldsymbol{h}_{1}&=\dfrac{1}{\sqrt{3}}\left(1,1,-1\right),\ \boldsymbol{h}_{2}=\dfrac{1}{\sqrt{3}}\left(1,-1,1\right),\ \boldsymbol{h}_{3}=\dfrac{1}{\sqrt{3}}\left(-1,1,1\right),\ \boldsymbol{h}_{4}=\dfrac{1}{\sqrt{3}}\left(-1,-1,-1\right).\nonumber
\end{align}
We introduce the dimensionless quantity $j_\mathrm{c}=J_\mathrm{c}/J$ and obtain the characteristic polynomial for the eigenvalues of the Hamiltonian \eqref{Ham_tetr},
\begin{equation}
    \!\!\!\!(2x\!-\!1)\!\left\{3(2x-1)^{2}(2x+3)-4j_\mathrm{c}^{2}(6x+1)\right\}^{3}\left[64j_\mathrm{c}^4(2x+1)^{2}+(4x^{2}-9)^{2}(2x-1)^{2}-4j_\mathrm{c}^{2}(80x^{4}-232x^{2}+117)\right]\!=\!0.\!\!\label{charact_pol}
\end{equation}
From this we can infer that at $j_\mathrm{c}=0$ the ground state with the energy $x=-3/2$ is fivefold degenerate because the term in $\{...\}$ provides a threefold degenerate root and the term in $[...]$ contains the twofold degenerate root at $x=-3/2$.
Let us consider first the branch arising from the term $\{...\}$.
At small $|j_\mathrm{c}|$, one finds 
\begin{equation}
    x_{1,2,3}=-\dfrac{3}{2}-\dfrac{1}{3}j_\mathrm{c}^{2}+\dfrac{1}{36}j_\mathrm{c}^{4}+\mathcal{O}(j_\mathrm{c}^{6}).\label{x123}
\end{equation}
The corresponding eigenstates are polarized states with a slight distortion for $j_\mathrm{c}\neq 0$ (see Fig.~\ref{Fig1}\textbf{a}-\textbf{c}). 
In the term $[...]$ we can perform a Taylor expansion for $x \approx -3/2$, and obtain the following roots,
\begin{eqnarray}
    & x_{4}=-\dfrac{3}{2}-\dfrac{1}{3}j_\mathrm{c}^{2}+\dfrac{\sqrt{29}}{18}j_\mathrm{c}^{3}-\dfrac{41}{108}j_\mathrm{c}^4+\mathcal{O}(j_\mathrm{c}^{5}),\nonumber\\
    & x_{5}=-\dfrac{3}{2}-\dfrac{1}{3}j_\mathrm{c}^{2}-\dfrac{\sqrt{29}}{18}j_\mathrm{c}^{3}-\dfrac{41}{108}j_\mathrm{c}^4+\mathcal{O}(j_\mathrm{c}^{5}).\label{x45}
\end{eqnarray}
Thus, the ground state becomes unique for any small nonzero $|j_\mathrm{c}|$, and the energy gap between the ground state and the first excited state is of the order $|j_\mathrm{c}^{3}|$.
Both eigenstates correspond to the quantum BP with slightly different spin lengths (see Fig.~\ref{Fig1}\textbf{d}, \textbf{e}).

\begin{figure*}[ht!]
\centering
\includegraphics[width=17.7cm]{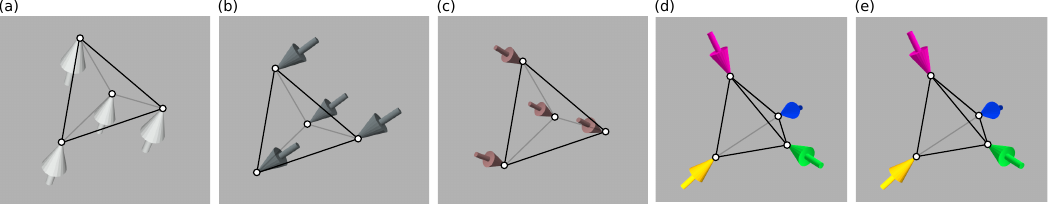}
\caption{~\small The five lowest-energy states. Panels (a)-(e) show the magnetic textures corresponding to the lowest-energy eigenstates calculated at $j_\mathrm{c}=0.01\sqrt{3}$. 
Panels (a)-(c) show the triplet of first excited states. Panels (d) and (e) are the GS and the second excited state, respectively, corresponding to the quantum BP. 
The spins in panels (d) and (e) have lengths of $0.00584957$ and $0.00569955$, respectively.  
} 
\label{Fig1}
\end{figure*}

\section{Wave function ansatz for $N=4$ case}

\begin{figure*}[ht!]
\centering
\includegraphics[width=18cm]{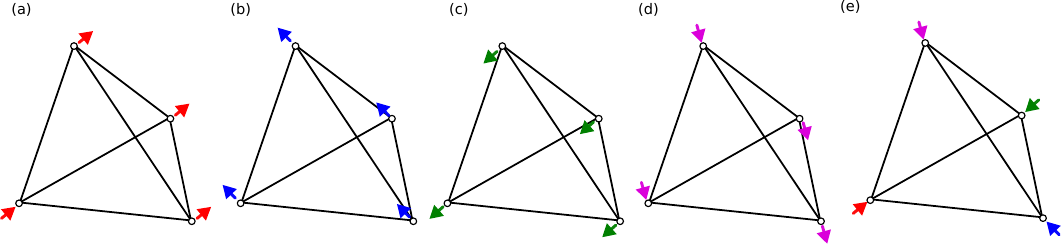}
\caption{~\small Construction of the tetragonal ansatz. Panels (a)-(d) show four uniformly magnetized spin textures and panel (e) contains the classical BP state. The directions of the arrows are given by the Zeeman fields $\bm{h}_{i}$ defined in Eq.~\eqref{Ham_tetr}.
} 
\label{Fig3}
\end{figure*}

As we have shown above, at $j_\mathrm{c}=0$, the state that we are interested in should correspond to zero total magnetization. 
In the tetrahedral case ($N=4$) such a state can be constructed as a linear superposition of four uniform states along the fields, $\bm{h}_i$ [see Fig.~\ref{Fig3}].
At the same time, in the limit of strong $j_\mathrm{c}$, the exchange interaction is negligibly small, and the state is the classical BP.
Taking this into account, we write the wave function ansatz in a so-called Anderson Towers form~\cite{tower},
\begin{eqnarray}
 & \ket{\psi_\mathrm{a}}=\alpha\ket{\psi_{0}}+i\beta \left(\ket{\psi_{1}}+\ket{\psi_{2}}+\ket{\psi_{3}}+\ket{\psi_{4}}\right)=\alpha\ket{\psi_{0}}+i\beta \ket{\tilde{\psi}},\label{psi_ansatz}
\end{eqnarray}
where $\alpha$ and $\beta$ are real parameters.
Defining the spinors
\begin{eqnarray}
    \ket{f_{1}}=e^{\mathrm{i}\pi/4}\left(\begin{array}{c}
        e^{-\mathrm{i}\phi}\sin\theta \\
        e^{\mathrm{i}\phi}\cos\theta  
    \end{array}\right),\,
    \ket{f_{2}}=e^{\mathrm{i}\pi/4}\left(\begin{array}{c}
        e^{\mathrm{i}\phi}\cos\theta \\
        e^{-\mathrm{i}\phi}\sin\theta  
    \end{array}\right),\,
    \ket{f_{3}}=e^{\mathrm{i}\pi/4}\left(\begin{array}{c}
        -e^{\mathrm{i}\phi}\cos\theta \\
        e^{-\mathrm{i}\phi}\sin\theta  
    \end{array}\right),\,
    \ket{f_{4}}=\left(\begin{array}{c}
        \mathrm{i} e^{\mathrm{i}\phi}\sin\theta \\
        e^{-\mathrm{i}\phi}\cos\theta  
    \end{array}\right),
\end{eqnarray}
where $\theta=\arccos(1/\sqrt{3})/2, \phi=\arcsin(1/(\sqrt{3}\sin 2\theta))/2$, the state $\ket{\psi_{i}}$ can be written as
\begin{eqnarray}
 && \ket{\psi_{i}}=\ket{f_{i}f_{i}f_{i}f_{i}},\ \forall i\in \{1,2,3,4\} , \,\, \mathrm{and} \,\, \ket{\psi_{0}}=\mathrm{i}\ket{f_{4}f_{3}f_{2}f_{1}}.
\end{eqnarray}

As one can notice, the state $\ket{\psi_{0}}$ is the ground state of a single shell Hamiltonian at
$j_c\rightarrow\infty$, while $\ket{\tilde{\psi}}$ is one of its possible ground states at $j_{c} \to 0$. 
The symmetry motivates such a choice for $\ket{\tilde{\psi}}$.

The fitting parameters $\alpha$, $\beta$ can be calculated as follows.
First of all, we notice that,
\begin{eqnarray}
& \braket{\psi_{i}|\psi_{j}}=\delta_{ij}+(1-\delta_{ij})/9,\nonumber\\
& \braket{\psi_{0}|\psi_{j}}=-\braket{\psi_{j}|\psi_{0}}=\dfrac{\mathrm{i}}{3\sqrt{3}},
\end{eqnarray}
Then, we can obtain
\begin{eqnarray}
 & \braket{\psi_\mathrm{a}\vert\psi_\mathrm{a}}=\alpha^{2}-\alpha\beta\dfrac{8}{3\sqrt{3}}+\dfrac{16}{3}\beta^{2}=1.\label{normalization}
\end{eqnarray}
Due to the spherical symmetry, the absolute values of all magnetization components are the same, so we can define the magnetization length as 
\begin{eqnarray}
     m=\vert\bra{\psi_\mathrm{a}}S_{x,1}\ket{\psi_\mathrm{a}}\vert/\sqrt{3}=\alpha^{2}-\dfrac{8}{3\sqrt{3}}\alpha\beta=1-\dfrac{16}{3}\beta^{2},\label{Magnetization}
\end{eqnarray}
The energy of this state is 
\begin{eqnarray}
E=\bra{\psi_\mathrm{a}}\mathcal{H}_{1}\ket{\psi_\mathrm{a}}=\dfrac{\alpha^{2}}{2}\left(J-4J_\mathrm{c}\right)-8J\beta^{2}+\dfrac{4}{\sqrt{3}}\alpha\beta\left(J+\dfrac{4}{3}J_\mathrm{c}\right).\label{En_ansatz}
\end{eqnarray}
The condition \eqref{normalization} can be satisfied by introducing new variables $\alpha^{\prime}, \beta^{\prime}$. These are defined as
\begin{eqnarray}
\alpha=\alpha^{\prime}\cos{\chi}-\beta^{\prime}\sin\chi,\qquad \beta=\beta^{\prime}\cos{\chi}+\alpha^{\prime}\sin\chi,
\end{eqnarray}
where the angle $\chi$ satisfies
\begin{equation}
    \dfrac{13}{3}\sin2\chi-\dfrac{8}{3\sqrt{3}}\cos2\chi=0 \quad\Longrightarrow\quad  \chi=\dfrac{1}{2}\arctan\dfrac{8}{13\sqrt{3}}\approx0.170691.\label{chi}
\end{equation}
In this case, the condition \eqref{normalization} writes as $A\alpha^{\prime2}+B\beta^{\prime2}=1$ where
\begin{align}
A&=\cos^{2}\chi-\dfrac{4}{3\sqrt{3}}\sin2\chi+\dfrac{16}{3}\sin^{2}\chi \notag \\
B&=\sin^{2}\chi+\dfrac{4}{3\sqrt{3}}\sin2\chi+\dfrac{16}{3}\cos^{2}\chi.
\end{align}
For $\chi$ defined in Eq.~\eqref{chi}, the parameters $A$ and $B$ are positive, so one can write
\begin{align}
\alpha&=\dfrac{1}{\sqrt{A}}\cos{\chi}\cos\xi-\dfrac{1}{\sqrt{B}}\sin\chi\sin\xi,\notag \\
\beta&=\dfrac{1}{\sqrt{B}}\cos{\chi}\sin\xi+\dfrac{1}{\sqrt{A}}\sin\chi\cos\xi, 
\end{align}
where the variable $\xi$ remains to be determined by minimizing the energy \eqref{En_ansatz}.
The energy minimum then corresponds to
\begin{equation}
\tan2\xi=4\sqrt{2}\dfrac{16 J_\mathrm{c}-3 J}{123 J-85 J_\mathrm{c}}\label{tan_xi_sol}
\end{equation}
Note that the right-hand side of Eq.~\eqref{tan_xi_sol} has a singularity at $J_\mathrm{c}=123 J/85$. Thus, the physical solution for $\xi$ can be written as
\begin{equation}
    \xi=\begin{cases}
        \dfrac{1}{2}\arctan\left[4\sqrt{2}\dfrac{16 J_\mathrm{c}-3 J}{123 J-85 J_\mathrm{c}}\right]+\dfrac{\pi}{2} & \text{for } 0<\dfrac{J_\mathrm{c}}{J}\leq\dfrac{123}{85},\\
        \dfrac{1}{2}\arctan\left[4\sqrt{2}\dfrac{16 J_\mathrm{c}-3 J}{123 J-85 J_\mathrm{c}}\right] & \text{for } \dfrac{J_\mathrm{c}}{J}>\dfrac{123}{85}.
    \end{cases}
\end{equation}
Using this ansatz, we can calculate the energy difference $\Delta E$ and the state overlap $\Delta \psi$ with the real ground state $\ket{\psi}$
\begin{eqnarray}
& \Delta{E}=\bra{\psi_\mathrm{a}}\mathcal{H}\ket{\psi_\mathrm{a}}-\bra{\psi}\mathcal{H}\ket{\psi}\geq0,\,\,\,\Delta\psi=1-|\braket{\psi_{a}|\psi}|.\label{Delta}
\end{eqnarray}
As expected, the errors decrease in the limits $J_\mathrm{c}/J\rightarrow0$ and $J_\mathrm{c}/J\rightarrow\infty$.

\begin{figure*}[ht!]
\centering
\includegraphics[width=17.7cm]{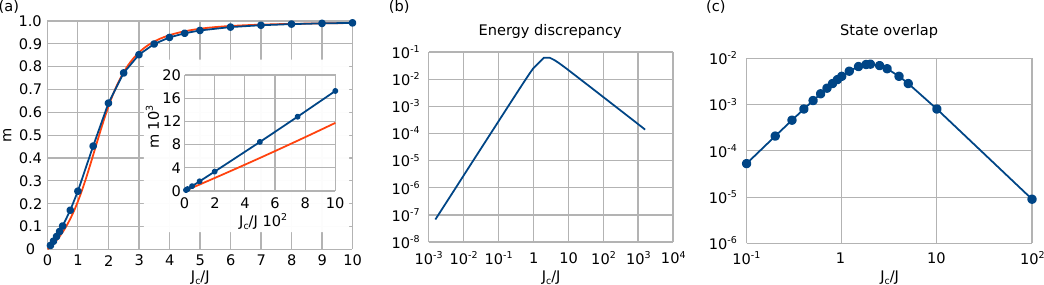}
\caption{~\small Comparison between numerical and analytical results.
Panel (a) shows the magnetization length for different coupling parameters $J_\mathrm{c}/J$. The blue curve is obtained numerically, the red one corresponds to Eq.~\eqref{Magnetization}.
Panels (b) and (c) show, respectively, the energy difference to the real ground state $\ket{\psi}$ and the overlap of the states $\ket{\psi_{\mathrm{a}}}$ and $\ket{\psi}$ according to Eq.~\eqref{Delta}. 
} 
\label{Fig4}
\end{figure*}

\section{Inter-shell interaction}

To motivate the choice of the multishell Hamiltonian [Eq.~(2) in the main text], we consider the following continuous energy functional for a spherical shell with volume $V_\mathrm{s}$,
\begin{equation}
\mathcal{E}_\mathrm{s}=\mathcal{A}\intop_{V_\mathrm{s}}\left[\left(\partial_x\bm{m}\right)^{2}+\left(\partial_y\bm{m}\right)^{2}+\left(\partial_z\bm{m}\right)^{2}\right]\mathrm{d}V.\label{int1}
\end{equation}
In spherical coordinates, it can be written as,
\begin{equation}
\mathcal{E}_\mathrm{s}=\mathcal{A}\intop_{0}^{2\pi}\intop_{0}^{\pi}\intop_{R}^{R+\delta r}\left[\left(\partial_r\bm{m}\right)^{2}+\dfrac{1}{r^{2}}\left(\partial_\theta\bm{m}\right)^{2}+\dfrac{1}{r^{2}\sin^{2}\theta}\left(\partial_\phi\bm{m}\right)^{2}\right]r^{2}\sin\theta\mathrm{d}r\mathrm{d}\theta\mathrm{d}\phi,
\end{equation}
where $R$ is the shell radius and $\delta r$ is its thickness.
This integral can be split into two parts, describing inter- and intra-shell interactions: 
\begin{equation} \mathcal{E}_\mathrm{s}=\mathcal{A}\sum_{i}\omega_{i}\intop_{R}^{R+\delta r}\left(\partial_r\bm{m}_{i}\right)^{2} r^{2}\mathrm{d}r + \mathcal{A}\delta r\intop_{0}^{2\pi}\intop_{0}^{\pi}\left[\left(\partial_\theta\bm{m}\right)^{2}+\dfrac{1}{\sin^{2}\theta}\left(\partial_\phi\bm{m}\right)^{2}\right]\sin\theta\mathrm{d}\theta\mathrm{d}\phi,\,\,\sum_{i}\omega_{i}=4\pi,\label{int3}
\end{equation}
where in the first integral, we substituted the integral over the spherical surface by discrete summation with some weights $\omega_{i}$ according to Ref.~\cite{Hesse}. In the second integral, we performed the integration over the radial variable. 
Assuming that each $\bm{m}_{i}$ remains constant on the interval $[R,R+\delta r]$, the first integral in Eq.~\eqref{int3} can be integrated over the radial coordinate,  
\begin{equation} \intop_{R}^{R+\delta r}\left(\partial_r\bm{m}_{i}\right)^{2}r^{2}\mathrm{d}r = \dfrac{1}{3}\left(\partial_r\bm{m}_{i}\right)^{2}r^{3}\Big|_{R}^{R+\delta r} = \left(\partial_r\bm{m}_{i}\right)^{2}R^{2}\delta r + \mathcal{O}(\delta r^{2}) = E_{0} - 2\bm{m}_{i,p}\cdot\bm{m}_{i,p+1}\left(\dfrac{R}{\delta r}\right)^{2}\delta r+\mathcal{O}(\delta r^{2}). \label{int4} 
\end{equation}
Here, we also substituted the radial derivative with a finite difference, leading to an inner product between magnetizations in $p$ and $p+1$ shells.
The shell radius is then given by $R=p\delta r$. Thus, the introduced $E_{0}=\left(|\bm{m}_{i,p}|^{2}+|\bm{m}_{i,p+1}|^{2}\right)p^{2}\delta r$ represents a constant contribution to the energy.

The second integral in \eqref{int3} represents a surface integral over the sphere, and it does not depend on the shell radius,
\begin{equation}
\intop_{0}^{2\pi}\intop_{0}^{\pi}\left[\left(\partial_{\theta}\bm{m}\right)^{2}+\dfrac{1}{\sin^{2}\theta}\left(\partial_{\phi}\bm{m}\right)^{2}\right]\sin\theta\mathrm{d}\theta\mathrm{d}\phi =\intop_{S}\left[\left(\nabla m_{x}\right)^{2}+\left(\nabla m_{y}\right)^{2}+\left(\nabla m_{z}\right)^{2}\right]\mathrm{d}S,\label{int5}
\end{equation}
where the $\nabla$ operator is taken in surface coordinates of the sphere $S$.
This integral can be approximated with a discrete sum corresponding to the uniform grid on the unit sphere,
\begin{equation}
\intop_{S}\left[\left(\nabla m_{x}\right)^{2}+\left(\nabla m_{y}\right)^{2}+\left(\nabla m_{z}\right)^{2}\right]\mathrm{d}S\approx e_{0}-2\sum_{\left\langle i,j\right\rangle}\omega_{i}\bm{m}_{i}\cdot\bm{m}_{j},\label{int6}
\end{equation}
where $e_{0}\propto N|\bm{m}|^{2}$ is a constant surface energy.
The values of the weights $\omega_{i}$ depend on how the points are distributed over the sphere. In the case of Platonic solids corresponding to uniform grids, one has a constant $\omega_{i}=\omega$.
Then, combining Eqs.~\eqref{int4} and \eqref{int5}, we get the atomistic version of the Hamiltonian \eqref{int1}, 
\begin{equation}
\mathcal{E}_\mathrm{s}\approx\mathcal{E}_{0}-2\mathcal{A}\omega\delta r\left[\sum_{\left\langle i,j\right\rangle}\bm{m}_{i}\cdot\bm{m}_{j}+ p^{2}\sum_{i}\bm{m}_{i,p}\cdot\bm{m}_{i,p+1}\right],\label{Ham_a_m}
\end{equation}
by neglecting terms proportional to $\delta r^{2}$.
Here, the constant energy $\mathcal{E}_{0}$ accounts for the contributions $E_{0}$ and $e_{0}$, which we omit in the main text in the multi-shell Hamiltonian.
The summation over the shells can be performed straightforwardly in Eq.~\eqref{Ham_a_m}. The exchange stiffness constant in the atomistic case is connected to the continuum one as $J=2\mathcal{A}\omega\delta r$.
Note that to get a strict equality in Eq.~\eqref{Ham_a_m}, one has to increase the number of spins $N$. At the same time, for $N>20$ it is impossible to construct a uniform grid for the sphere. That means in the limit $N\rightarrow\infty$, one must deal with non-uniform grids, which in turn will require utilizing distinct values for $\omega_{i}$ as explained in Ref.~\cite{Hesse}.

\section{Asymptotic behavior of the quantum Bloch point}
The energy for a quantum BP in a ball of radius $R$ provided in the main text has the form:
\begin{equation}
    \mathcal{E}_\mathrm{BP}=8\pi\mathcal{A}R + 4\pi\mathcal{A}\intop_{0}^{R}\left[\left(\psi^{\prime}\right)^{2}+\left(\!\dfrac{\kappa}{\mathcal{A}}\!-\!\dfrac{2}{r^{2}}\right)\sin^{2}\psi\right]r^{2}\mathrm{d}r.\label{BP_in_ball}
\end{equation}
Then the Euler-Lagrange equation following from $\delta \mathcal{E}_\mathrm{BP}=0$ is of the form:
\begin{equation}
r^{2}\psi^{\prime\prime}+2r\psi^{\prime}+\left(2-\left(\dfrac{r}{r^{*}}\right)^{2}\right)\dfrac{\sin2\psi}{2}=0.\label{psi_eq1}
\end{equation}
In the asymptotic limit $r\gg1$, we have $\psi\rightarrow0$ and the linearized equation can be written as
\begin{equation}
r^{2}\psi^{\prime\prime}+2r\psi^{\prime}+\left(2-\left(\dfrac{r}{r^{*}}\right)^{2}\right)\psi=0.\label{psi_eq2}
\end{equation}
% %
% %
In the limit $r\rightarrow\infty$, we may neglect the term $2\psi$ from the previous differential equation and, using $g=r\psi$, reduce it to $g''-g/r^{*2}=0$ with solution $g=a e^{-r/r^*}$.

On the other hand, near the center of the quantum BP at $r\ll1$, we have $\psi \approx \pi/2-\chi$ with small $\chi$, so the linearized equation is a order-$1$ Bessel equation
\begin{equation}
r^{2}\chi^{\prime\prime}+2r\chi^{\prime}-\left(2-\left(\dfrac{r}{r^{*}}\right)^{2}\right)\chi=0.\label{psi_eq3}
\end{equation}
The two linearly independent solutions to this equation are spherical Bessel functions.
Using Rayleigh's formulae, we get
\begin{equation}
    \chi(r)=\dfrac{r^{*}}{r}\left(b_{1}\left(\sin\dfrac{r}{r^{*}}+\dfrac{r^{*}}{r}\cos\dfrac{r}{r^{*}} \right)+b_{2}\left(\cos\dfrac{r}{r^{*}}-\dfrac{r^{*}}{r}\sin\dfrac{r}{r^{*}} \right)\right),
\end{equation}
where $b_{1}, b_{2}$ are arbitrary constants and we have to set $b_{1}=0$ to keep only convergent part of the solution.
Thus, the magnetization $\bm{n}(r)=-\bm{e}_{\mathrm{r}}\cos\psi $ in the vicinity of the quantum BP core is described as
\begin{equation}
    \bm{n}(x,y,z)\approx -\bm{e}_{\mathrm{r}} \chi  =-b_{2}\dfrac{r^{*}}{r}\left(\cos\dfrac{r}{r^{*}}-\dfrac{r^{*}}{r}\sin\dfrac{r}{r^{*}} \right)\bm{e}_{\mathrm{r}}.
\end{equation}
The effective field is given by,
\begin{equation}
\bm{b}_\mathrm{eff}\propto\triangle\bm{n}=\triangle\left(-\dfrac{b_{2}}{r}\left[\cos\dfrac{r}{r^{*}}-\dfrac{r^{*}}{r}\sin\dfrac{r}{r^{*}} \right]\bm{e}_{\mathrm{r}}\right)= r\left[\dfrac{b_{2}}{3(r^{*})^{3}}+\mathcal{O}(r^{2})\right]\bm{e}_{\mathrm{r}}.
\end{equation}
At the same time, for the classical BP one has,
\begin{equation}
\bm{b}_\mathrm{eff}\propto\triangle\bm{m}=\dfrac{2}{r^{2}}\bm{e}_{\mathrm{r}}.
\end{equation}

\section{Ansatz for the quantum Bloch point profile}
In terms of the dimensionless coordinate $\rho=r/r^{*}$, where $r^{*}=\sqrt{\mathcal{A}/\kappa}$, the energy of the quantum BP described by $\psi(\rho)$ has the form
\begin{eqnarray}
 & \mathcal{E}_\mathrm{BP}=8\pi\mathcal{A}R+4\pi\mathcal{A}r^{*} \mathcal{I}\left(\dfrac{R}{r^{*}}\right)=8\pi\mathcal{A}R+ 4\pi\mathcal{A}r^{*}\displaystyle\intop_{0}^{R/r^{*}}\left[\left(\dfrac{\mathrm{d}\psi}{\mathrm{d}\rho}\right)^2+\left(1-\dfrac{2}{\rho^{2}}\right)\sin^{2}\psi\right]\rho^{ 2}\mathrm{d}\rho,
 \label{BP_energy_dim}
\end{eqnarray}
Employing the ansatz $\psi_\mathrm{a}(\rho)=\arccos\tanh(c\rho)$ we get the following integral $\mathcal{I}$ that has to be calculated:
\begin{eqnarray}
    && \mathcal{I}\left(\dfrac{R}{r^{*}}\right)=\displaystyle\intop_{0}^{R/r^{*}}\dfrac{(1+c^{2})\rho^{2}-2}{\cosh^{2}c \rho}\mathrm{d}\rho.
    \label{integral_ansatz}
\end{eqnarray}
Performing the integration leads to
\begin{eqnarray}
    &&\!\!\!\!\!\!\!\!\!\!\!\!\!\!\mathcal{I}\left(\!\dfrac{R}{r^{*}}\!\right)\!=\!\dfrac{(1+c^2)\pi^2}{12c^3}\!-\dfrac{1+c^2}{c^2}\left(\dfrac{cR^{2}}{r^{* 2}}\left(1-\tanh\dfrac{cR}{r^{*}}\right)+\dfrac{2R}{r^{*}}\ln\left[1+e^{-2 c R/r^{*}}\right]\!-\!\dfrac{1}{c}\mathrm{Li}_{2}\left(-e^{-2 c R/r^{*}}\right)\right)-\dfrac{2}{c}\tanh\dfrac{cR}{r^{*}},
    \label{IRxi}
\end{eqnarray}
where $\mathrm{Li}_{2}$ is the polylogarithm function.
In the limit $R/r^{*}\rightarrow\infty$, the expression~\eqref{IRxi} can be simplified to
\begin{equation}
    \mathcal{I}(\infty)=\dfrac{(1+c^2)\pi^2}{12c^3}-\dfrac{2}{c},
    \label{IRxi_limit}
\end{equation}
and we can find the value of $c^{*}$ which minimizes Eq.~\eqref{IRxi_limit} from the equation,
\begin{equation}
   \dfrac{2}{c^{2}}-\dfrac{3+c^{2}}{12c^{4}}\pi^2=0.
    \label{a_eq}
\end{equation}
One thus has $c^{*}=\pi\sqrt{3/(24-\pi^2)}$, where the second solution can be omitted due to the condition $\psi(\infty)=0$.
The value of $\mathcal{I}$ at this value for $c^{*}$ is $-\pi^2/(6 c^{* 3})\approx-0.\mathbf{54}23$.
The exact value of this integral was found numerically by minimizing \eqref{BP_energy_dim} and it equals $-0.\mathbf{54}39$, so the suggested ansatz gives indeed a good estimation for the hedgehog profile at $R/r^{*}\rightarrow \infty$. 

\section{Details of micromagnetic simulations}
Using the four-dimensional vector $(n_{x},n_{y},n_{z},b)$ with the constraint $n_{x}^{2}+n_{y}^{2}+n_{z}^2+b^{2}=1$, we performed the energy minimization with the nonlinear conjugated gradients method as described in Ref.~\cite{Rybakov_thesis}.  
The presented approach suggests parametrizing the magnetization by stereographic projections from two poles to avoid slowing down the algorithm.
In our case, the method can be directly generalized, and we write
\begin{equation}
    n_{i}=\dfrac{2\gamma_{i}}{1+\gamma^{2}},\quad b=\gamma_{0}\dfrac{1-\gamma^{2}}{1+\gamma^{2}}, \quad i\in \{x,y,z\},\label{stereographic}
\end{equation}
where $\gamma^{2}=\gamma_{x}^2+\gamma_{y}^2+\gamma_{z}^2$, and $\gamma_{0}=+1$ if $b\geq0$ and $\gamma_{0}=-1$ if $b<0$. 
In this case, we need to solve the unconstrained optimization problem for the order parameter $(\gamma_{x},\gamma_{y},\gamma_{z})\in\mathbb{R}^{3}$ and $\gamma_{0} \in \{-1,1 \}$ defined for each spin.

\section{Polarized magnetic small-angle neutron scattering cross section}

\begin{figure}[tb!]
\centering
\resizebox{0.50\columnwidth}{!}{\includegraphics{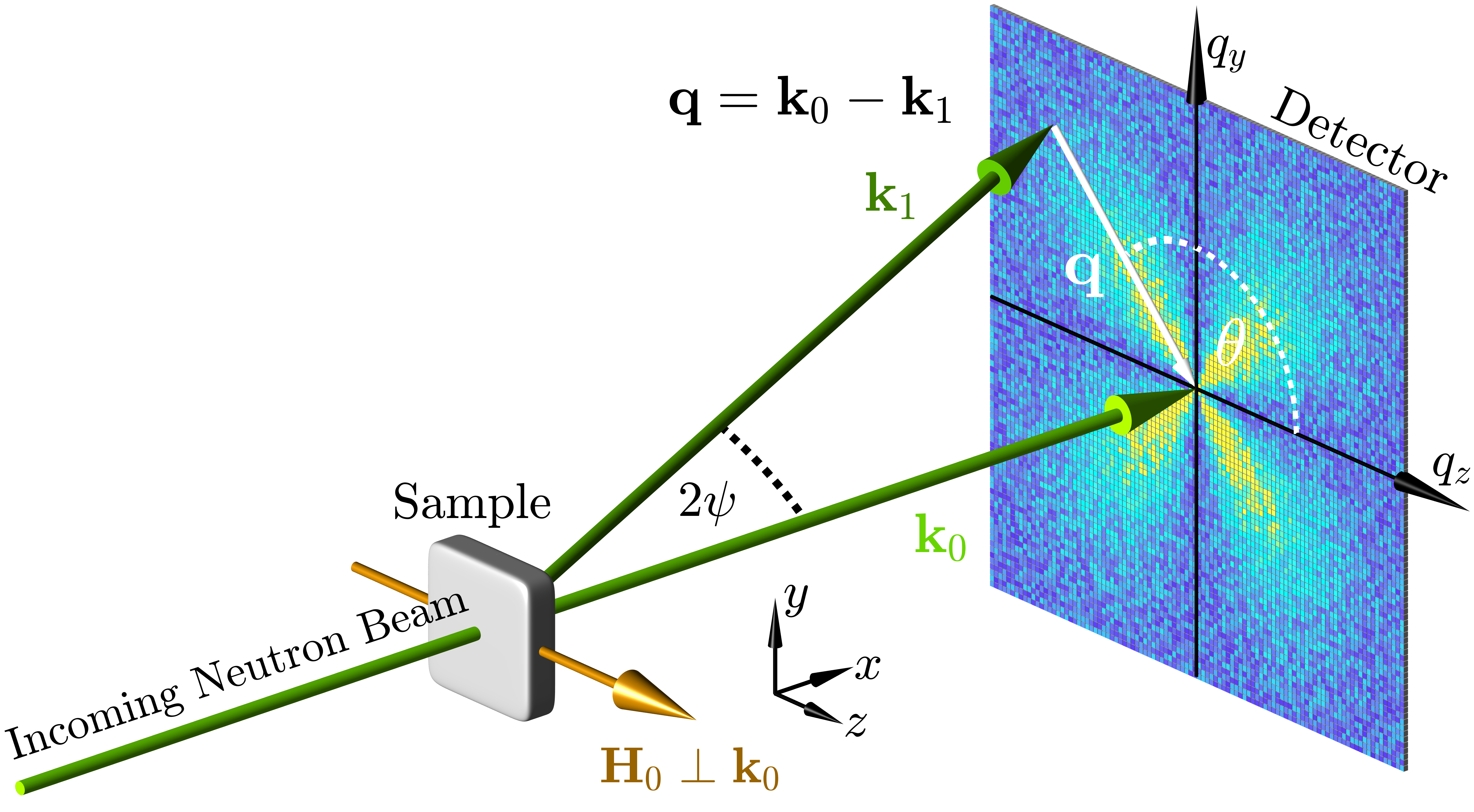}}
\caption{Sketch of the scattering geometry assumed in the micromagnetic simulations. The neutron optical elements (polarizer, spin flipper, analyzer) that are required to measure the spin-flip SANS cross section and the chiral function are not drawn. The applied magnetic field $\mathbf{H}_0 \parallel \mathbf{e}_z$ is perpendicular to the wave vector $\mathbf{k}_0 \parallel \mathbf{e}_x$ of the incident neutron beam ($\mathbf{H}_0 \perp \mathbf{k}_0$). The momentum-transfer or scattering vector $\mathbf{q}$ is defined as the difference between $\mathbf{k}_0$ and $\mathbf{k}_1$, i.e., $\mathbf{q} = \mathbf{k}_0 - \mathbf{k}_1$. SANS is usually implemented as elastic scattering ($k_0 = k_1 = 2\pi / \lambda$), and the component of $\mathbf{q}$ along the incident neutron beam, here $q_x$, is much smaller than the other two components so that $\mathbf{q} \cong \{ 0, q_y, q_z \} = q \{ 0, \sin\theta, \cos\theta \}$. This demonstrates that SANS probes predominantly correlations in the plane perpendicular to the incident beam. The angle $\theta = \angle(\mathbf{q}, \mathbf{H}_0)$ is used to describe the angular anisotropy of the recorded scattering pattern on the two-dimensional position-sensitive detector. For elastic scattering, the magnitude of $\mathbf{q}$ is given by $q = (4\pi / \lambda) \sin(\psi/2)$, where $\lambda$ denotes the mean wavelength of the neutrons and $\psi$ is the scattering angle.}
\label{sanssetup}
\end{figure}

In the context of magnetic SANS experiments, we are interested in the elastic differential spin-flip scattering cross section and the related so-called chiral function, which are experimentally accessible.
For the most common scattering geometry in SANS experiments (see Fig.~\ref{sanssetup}), where the applied magnetic field $\mathbf{H}_0 \parallel \mathbf{e}_z$ is perpendicular to the wave vector $\mathbf{k}_0 \parallel \mathbf{e}_x$ of the incident neutrons, the two spin-flip SANS cross sections $d \Sigma^{+-}_{\mathrm{sf}} / d \Omega$ and $d \Sigma^{-+}_{\mathrm{sf}} / d \Omega$ can be written as~\cite{Michels}:
\begin{eqnarray}
 \label{eq:equation1a}
\frac{d \Sigma^{+-}_{\mathrm{sf}}}{d \Omega} &=& \frac{8 \pi^3}{V} b_{\mathrm{H}}^2 \left( |\widetilde{M}_x|^2 + |\widetilde{M}_y|^2 \cos^4\theta + |\widetilde{M}_z|^2 \sin^2\theta \cos^2\theta \right. \\ &&\left. - (\widetilde{M}_y \widetilde{M}_z^{\ast} + \widetilde{M}_y^{\ast} \widetilde{M}_z) \sin\theta \cos^3\theta - i \chi \right), \nonumber \\
\frac{d \Sigma^{-+}_{\mathrm{sf}}}{d \Omega} &=& \frac{8 \pi^3}{V} b_{\mathrm{H}}^2 \left( |\widetilde{M}_x|^2 + |\widetilde{M}_y|^2 \cos^4\theta + |\widetilde{M}_z|^2 \sin^2\theta \cos^2\theta \right. \\ &&\left. - (\widetilde{M}_y \widetilde{M}_z^{\ast} + \widetilde{M}_y^{\ast} \widetilde{M}_z) \sin\theta \cos^3\theta + i \chi \right) \nonumber .
 \label{eq:equation1b}
\end{eqnarray}
The superscripts refer to the neutron-spin orientation, parallel ($+$) or antiparallel ($-$), relative to the direction of $\mathbf{H}_0$, $V$ is the scattering volume, $b_{\mathrm{H}} = 2.91 \times 10^8 \, \mathrm{A}^{-1}\mathrm{m}^{-1}$ is the magnetic scattering length in the small-angle regime (the atomic magnetic form factor is approximated by $1$ due to the forward scattering), $\widetilde{\mathbf{M}}(\mathbf{q}) = \{ \widetilde{M}_x(\mathbf{q}), \widetilde{M}_y(\mathbf{q}), \widetilde{M}_z(\mathbf{q}) \}$ represents the Fourier transform of the magnetization vector field $\mathbf{M}(\mathbf{r}) = \{ M_x(\mathbf{r}), M_y(\mathbf{r}), M_z(\mathbf{r}) \}$, $\theta$ is the angle between $\mathbf{q}$ and $\mathbf{H}_0$, the asterisk ``$*$'' marks the complex-conjugated quantity, $i^2 = -1$, and $\chi = \chi(\mathbf{q})$ is the chiral function. The latter quantity is obtained from the difference between the two spin-flip SANS cross sections, according to~\cite{Michels}:
\begin{eqnarray}
\label{chiralfunc1}
-i K \chi(\mathbf{q}) = \frac{1}{2} \left( \frac{d \Sigma^{+-}_{\mathrm{sf}}}{d \Omega} - \frac{d \Sigma^{-+}_{\mathrm{sf}}}{d \Omega} \right)
= -i K \left[ (\widetilde{M}_x \widetilde{M}_y^{\ast} - \widetilde{M}_x^{\ast} \widetilde{M}_y) \cos^2\theta - (\widetilde{M}_x \widetilde{M}_z^{\ast} - \widetilde{M}_x^{\ast} \widetilde{M}_z) \sin\theta \cos\theta \right],
\end{eqnarray}
where $K = \frac{8 \pi^3}{V} b_{\mathrm{H}}^2$. 
Besides the difference between $d \Sigma^{+-}_{\mathrm{sf}} / d \Omega$ and $d \Sigma^{-+}_{\mathrm{sf}} / d \Omega$, one can also consider the following sum:
\begin{eqnarray}
\frac{d \Sigma_{\mathrm{sf}}}{d \Omega} = \frac{1}{2} \left( \frac{d \Sigma^{+-}_{\mathrm{sf}}}{d \Omega} + \frac{d \Sigma^{-+}_{\mathrm{sf}}}{d \Omega} \right) = K \left( |\widetilde{M}_x|^2 + |\widetilde{M}_y|^2 \cos^4\theta + |\widetilde{M}_z|^2 \sin^2\theta \cos^2\theta - (\widetilde{M}_y \widetilde{M}_z^{\ast} + \widetilde{M}_y^{\ast} \widetilde{M}_z) \sin\theta \cos^3\theta \right).\label{chiral444}
\end{eqnarray}
The quantity $d \Sigma_{\mathrm{sf}} / d \Omega$ is the polarization-independent spin-flip SANS cross section. 
The following symmetry relations hold for $d \Sigma_{\mathrm{sf}} / d \Omega$ (even under spatial inversion of $\mathbf{q}$) and $-i K \chi$ (odd under spatial inversion of $\mathbf{q}$):
\begin{equation}
\frac{d \Sigma_{\mathrm{sf}}}{d \Omega}(\mathbf{q}) = \frac{d \Sigma_{\mathrm{sf}}}{d \Omega}(-\mathbf{q}) .
\label{chiral4}
\end{equation}
\begin{equation}
i K \chi(\mathbf{q}) = - i K \chi(-\mathbf{q}) .
 \label{chiral3}
\end{equation}
We emphasize that the chiral function vanishes at complete magnetic saturation ($M_x^{H_0 \rightarrow \infty} = M_y^{H_0 \rightarrow \infty} = 0$) and for purely real or for purely imaginary magnetization Fourier components $\widetilde{M}_{x,y,z}$.

It is often convenient to average two-dimensional SANS data $f(\mathbf{q}) = f(q_y, q_z) = f(q, \theta)$, where $f$ either stands for $d \Sigma_{\mathrm{sf}} / d \Omega$ or for $-i K \chi$, along certain directions in $\mathbf{q}$~space, e.g.\ parallel ($\theta = 0$) or perpendicular ($\theta = \pi/2$) to the applied magnetic field, or even over the full angular $\theta$~range. Here, we focus on $2\pi$~azimuthally-averaged SANS data
\begin{equation}
\label{aziaverage}
I_{\mathrm{sf}}(q) = \frac{1}{2\pi} \int_0^{2\pi} f(q,\theta) \, d\theta ,
\end{equation}
which allows for the computation of the pair-distance distribution function $p_{\mathrm{sf}}(r)$ according to
\begin{equation}
\label{pvonreqintegral}
p_{\mathrm{sf}}(r) = r \int\limits_0^{\infty} I_{\mathrm{sf}}(q) \sin(qr) q dq .
\end{equation}
This Fourier transform relates to the distribution of real-space distances between volume elements inside the particle weighted by the excess scattering-length density distribution, as discussed in detail in the review by Svergun and Koch~\cite{Svergun}.
As a reference for spherical particles with a nonuniform magnetization distribution, we specify here the $p_{\mathrm{sf}}(r)$ of a uniformly magnetized sphere, which for $r \leq D = 2R$ equals:
\begin{equation}
\label{pvonreq}
p_{\mathrm{sf}}(r) \propto r^2 \left( 1 - \frac{3r}{4R} + \frac{r^3}{16 R^3} \right) .
\end{equation}

\section{Bloch point experimental signature}
\subsection{Autocorrelation functions}
Considering the BP texture stabilized in a nanosphere, we can calculate the autocorrelation function,
\begin{equation}
    c(\bm{a}) = \intop \bm{M}\left(\bm{r}-\dfrac{\bm{a}}{2}\right)\cdot\bm{M}\left(\bm{r}+\dfrac{\bm{a}}{2}\right)\mathrm{d}V,
\end{equation}
where the total sphere displacement is given by vector $\bm{a}$, and the integration is over volume given by the intersection of the two spheres.
Without loss of generality, we use $\bm{a}=(0,0,a)$.
For a sphere of radius $R=1$ and employing the quantum BP ansatz
\begin{equation}
    \bm{M}(\bm{r})=\bm{n}(\bm{r})=\dfrac{(y,-x,-z)}{R},
\end{equation}
we obtain
\begin{equation}
    c_{1}(\bm{a}) = \intop \left(x^{2}+y^{2}+z^{2}-\dfrac{a^{2}}{4}\right)\mathrm{d}V=\dfrac{4\pi}{5}\left(1-\dfrac{5}{2}\dfrac{a}{2}+\dfrac{5}{2}\left(\dfrac{a}{2}\right)^{3}-\left(\dfrac{a}{2}\right)^{5}\right).
\end{equation}

On the other hand, for the case of a classical BP with the profile
\begin{equation}
    \bm{M}(\bm{r})=\bm{m}(\bm{r})=\dfrac{(y,-x,-z)}{\sqrt{x^{2}+y^{2}+z^{2}}},
\end{equation}
we obtain
\begin{equation}
    c_{2}(\bm{a}) = \intop \dfrac{x^{2}+y^{2}+z^{2}-a^{2}/4}{\sqrt{\left(x^{2}+y^{2}+z^{2}+a^{2}/4\right)^{2}-a^{2}z^{2}}}\mathrm{d}V=\dfrac{4\pi}{3}\begin{cases}
        1-\dfrac{3 a}{4}-a^{2}+\dfrac{3a^{3}}{4}, &\text{for } 0\leq a\leq 1,\\
        -1+\dfrac{1}{a}-\dfrac{3 a}{4}+a^{2}-\dfrac{a^{3}}{4}, & \text{for }1< a\leq 2.
    \end{cases}
    \label{eq:autocorrelation_classical}
\end{equation}

The resulting functions $c_{1,2}$ are shown in Fig.~1 of the main text
In the classical BP case, we have $c_{2}=0$ at $a=1$ and $a=2$. 
The former zero is due to the BP's spherical symmetry, and the latter corresponds to zero overlap of the spheres.
In contrast, in the quantum BP case, one finds zeros of $c_{1}$ at $a\approx1.07$ and $a=2$.
The change in the value of the first root is due to the possibility of spin length variation in a quantum case. 
Thus, this characteristic feature can be used to distinguish classical and quantum BPs.

\subsection{SANS cross section for the quantum BP}
Taking into account that
\begin{equation}
    \intop e^{-i \bm{q}\cdot\bm{r}}\mathrm{d}V=\dfrac{J_{3/2}(q)}{q^{3/2}}=4\pi\dfrac{\sin q-q\cos q}{q^{3}},
\end{equation}
where $q=\sqrt{q_{x}^{2}+q_{y}^{2}+q_{z}^{2}}$ we can obtain
\begin{equation}
\intop r_{j} e^{-i \bm{q}\cdot\bm{r}}\mathrm{d}V=4\pi i q_{j}\dfrac{3q\cos q+\left(q^{2}-3\right)\sin q}{q^{5}},
\end{equation}
for $j\in\{x,y,z\}$. Then, the Fourier transform of the quantum BP becomes
\begin{equation}
    \widetilde{\bm{n}}(\bm{q})=4\pi i(q_{y},-q_{x},-q_{z}) \dfrac{3q\cos q+\left(q^{2}-3\right)\sin q}{q^{5}}.
\end{equation}
The spin-flip SANS cross section Eq.~\eqref{chiral444} then takes the form
\begin{equation}
    \dfrac{\mathrm{d} \Sigma_{\mathrm{sf}}}{\mathrm{d} \Omega}(\mathbf{q})=16\pi^{2}K \left(1  +  \cos^{4}\theta  \right)\dfrac{\left(3q\cos q+\left(q^{2}-3\right)\sin q\right)^{2}}{q^{8}}\sin^{2}\theta .
\end{equation}
The azimuthally-averaged SANS cross section Eq.~\eqref{aziaverage} is given by
\begin{equation}
I_{\mathrm{sf}}(q) = 9\pi^{2}K \dfrac{\left(3q\cos q+\left(q^{2}-3\right)\sin q\right)^{2}}{q^{8}} ,
\end{equation}
from which the pair-distance distribution function Eq.~\eqref{pvonreqintegral} can be obtained,
\begin{equation}
p_{\mathrm{sf}}(r) \propto r^2 \left( 1 - \dfrac{5}{2}\dfrac{r}{2R} + \dfrac{5}{2}\dfrac{r^{3}}{(2R)^{3}} - \dfrac{r^{5}}{(2R)^{5}} \right).
\end{equation}

\subsection{SANS cross section for the classical BP}

We can evaluate the following integral
\begin{equation}
    \intop \dfrac{e^{-i \bm{q}\cdot\bm{r}}}{|\bm{r}|}\mathrm{d}V
    =
    \intop{\rm dr}{\rm d\theta}{\rm d\phi}\, r\sin\theta {\rm e}^{- iqr\cos\theta}
    =
    \frac{4 \pi (1-\cos q)}{q^2},
\end{equation}
from which we then obtain the components
\begin{equation}
\intop \dfrac{r_{j}e^{-i \bm{q}\cdot\bm{r}}}{|\bm{r}|}\mathrm{d}V
=
4\pi i q_j \frac{q\sin q + 2\cos q - 2}{q^4},
\end{equation}
for $j\in\{x,y,z\}$. The Fourier transform of a classical Bloch profile $\bm m(\bm r)$ results in
\begin{equation}
    \widetilde{\bm{m}}(\bm{q})
    =
    4\pi i(q_{y},-q_{x},-q_{z}) \frac{q\sin q + 2 \cos q - 2}{q^4}.
\end{equation}
Then, the spin-flip SANS cross section Eq.~\eqref{chiral444} is
\begin{equation}
    \dfrac{\mathrm{d} \Sigma_{\mathrm{sf}}}{\mathrm{d} \Omega}(\mathbf{q})
    =K
    \frac{16 \pi^2 \sin^2\theta \left(\cos^4\theta + 1\right)\left(q\sin q + 2\cos q - 2\right)^2}{q^6}
    ,
\end{equation}
and the azimuthally-averaged SANS data becomes
\begin{equation}
I_{\mathrm{sf}}(q) = K\dfrac{9\pi^2 (q\sin q + 2\cos q - 2)^2}{q^6}.
\label{eq:isf_classical}
\end{equation}
We can again derive the pair-distance distribution function via Eq.~\eqref{pvonreqintegral}, and find
\begin{equation}
    p_{\mathrm{sf}}(r) \propto r^{2} \left(-\dfrac{\left(\dfrac {r^{2}}{R^{2}}-1\right)\left(2\left(\dfrac {r}{R}-1\right)^3-\left(\dfrac {r^{2}}{R^{2}}-2\right)\left|\dfrac{r}{R}-1\right|\right)}{4\dfrac{r}{R}\left|\dfrac{r}{R}-1\right|}
        \right)
    ,\ 0\leq r/R\leq 2 
    .
\end{equation}
The expression in brackets shows the same scaling as the autocorrelation function in Eq.~\eqref{eq:autocorrelation_classical}, i.e., $p_{\mathrm{sf}}(aR) \sim c_2 a^2$.

\twocolumngrid


\begin{thebibliography}{99}
	\twocolumngrid
	
	\bibitem{Feldtkeller_1965}
	E. Feldtkeller, Mikromagnetisch stetige und unstetige Magnetisierungskonfigurationen,
	Z. Angew. Phys \textbf{19}, 530 -- 536, (1965).
	
	\bibitem{Doring_1968}
	D\"oring, W., Point Singularities in Micromagnetism, \href{http://link.aip.org/link/?JAP/39/1006/1}{J. Appl. Phys. \textbf{39},
		1006-1007, (1968).}
	
	\bibitem{Malozemoff_79}
	A.~P. Malozemoff and J.~C. Slonczewski, 
	\textit{Magnetic Domain Walls in Bubble Materials} (Academic Press, New York, 1979).
	
	\bibitem{Fruchart}
	S. Da Col, S. Jamet, N. Rougemaille, A. Locatelli, T. O. Mentes, B. Santos Burgos, R. Afid, M. Darques, L. Cagnon, J. C. Toussaint, and O. Fruchart, Observation of Bloch-point domain walls in cylindrical magnetic nanowires,
	\href{https://doi.org/10.1103/PhysRevB.89.180405}{Phys. Rev. B \textbf{89}, 180405(R) (2014).}
	
	\bibitem{Rybakov_2015}
	F. N. Rybakov, A. B. Borisov, S. Blügel, and N. S. Kiselev, New type of
	stable particlelike states in chiral magnets, Phys. Rev. Lett. \textbf{115}, 117201 (2015).
	
	\bibitem{Ran}
	K. Ran, Y. Liu, Y. Guang, D. M. Burn, Gerrit van der Laan, T. Hesjedal, H.  Du, G. Yu, and S. Zhang, Creation of a Chiral Bobber Lattice in Helimagnet-Multilayer Heterostructures, 
	\href{https://doi.org/10.1103/PhysRevLett.126.017204}{Phys. Rev. Lett. \textbf{126}, 017204 (2021).}
	
	\bibitem{globule}
	M\"uller, G. P., Rybakov, F. N., Jonsson, H., Bl\"ugel, S., and Kiselev, N. S.  Coupled quasimonopoles
	in chiral magnets.  \href{doi:10.1103/physrevb.101.184405}{Physical Review B \textbf{101} (2020).}
	
	\bibitem{Azhar}
	Maria Azhar, Volodymyr P. Kravchuk, and Markus Garst, Screw Dislocations in Chiral Magnets,
	\href{https://doi.org/10.1103/PhysRevLett.128.157204}{Phys. Rev. Lett. \textbf{128}, 157204 (2022).}
	
	\bibitem{hopfion}
	Fengshan Zheng, Nikolai S. Kiselev, Filipp N. Rybakov, Luyan Yang, Wen Shi, Stefan Blügel and Rafal E. Dunin-Borkowski, Hopfion rings in a cubic chiral magnet,  
	\href{https://doi.org/10.1038/s41586-023-06658-5}{Nature \textbf{623}, 718–723 (2023).} 
	
	\bibitem{Nagaosa}
	Yizhou Liu and Naoto Nagaosa, Current-Induced Creation of Topological Vortex Rings in a Magnetic Nanocylinder, 
	\href{https://doi.org/10.1103/PhysRevLett.132.126701}{Phys. Rev. Lett. \textbf{132}, 126701 (2024).}
	
	
	
	
	\bibitem{Beg}
	Beg, M., Pepper, R.A., Cortés-Ortuño, D. et al. Stable and manipulable Bloch point, \href{https://doi.org/10.1038/s41598-019-44462-2}{Sci Rep \textbf{9}, 7959 (2019).} 
	
	\bibitem{Fischer}
	Mi-Young Im, Hee-Sung Han, Min-Seung Jung, Young-Sang Yu, Sooseok Lee, Seongsoo Yoon, Weilun Chao, Peter Fischer, Jung-Il Hong and Ki-Suk Lee, 
	Dynamics of the Bloch point in an asymmetric permalloy disk. \href{https://doi.org/10.1038/s41467-019-08327-6}{Nat Commun \textbf{10}, 593 (2019).} 
	
	\bibitem{Guslienko}
	F.~Tejo, J.~A.~F.~Fernandez-Roldan, K.~Guslienko, R.~M.~Otxoa and O.~Chubykalo-Fesenko, Giant supermagnonic Bloch point velocities in cylindrical ferromagnetic nanowires, Nanoscale, (2024), DOI:
	10.1039/D3NR05013K.
	
	\bibitem{Aliev}
	Carlos Sánchez, Diego Caso, and Farkhad G. Aliev,
	Artificial Neuron Based on the Bloch-Point Domain Wall in Ferromagnetic Nanowires,
	\href{https://doi.org/10.3390/ma17102425}{Materials \textbf{17}(10), 2425, (2024).}
	
	
	\bibitem{Kwon}
	Se Kwon Kim and Oleg Tchernyshyov, Pinning of a Bloch point by an atomic lattice, \href{https://doi.org/10.1103/PhysRevB.88.174402}{Phys. Rev. B \textbf{88}, 174402, (2013).}
	
	\bibitem{Gong}
	Zizhao Gong, Jin Tang, Sergey S. Pershoguba, Zongkai Xie, Rui Sun, Yang Li, Xu Yang, Jianan Liu, Wei Zhang, Xiangqun Zhang, Wei He, Haifeng Du, Jiadong Zang, and Zhao-hua Cheng, Current-induced dynamics and tunable spectra of a magnetic chiral bobber, 
	\href{https://doi.org/10.1103/PhysRevB.104.L100412}{Phys. Rev. B \textbf{104}, L100412, (2021).}
	
	\bibitem{Hertel}
	Christian Andreas, Attila Kákay, and Riccardo Hertel, Multiscale and multimodel simulation of Bloch-point dynamics, \href{https://doi.org/10.1103/PhysRevB.89.134403}
	{Phys. Rev. B \textbf{89}, 134403 (2014).}
	
	\bibitem{Verga}
	Elías, R.G., Verga, A. Magnetization structure of a Bloch point singularity. \href{https://doi.org/10.1140/epjb/e2011-20146-6}{Eur. Phys. J. B \textbf{82}, 159–166 (2011).} 
	
	\bibitem{Chubykalo}
	K. M. Lebecki, D. Hinzke, U. Nowak, and O. Chubykalo-Fesenko, Key role of temperature in ferromagnetic Bloch point simulations, \href{https://doi.org/10.1103/PhysRevB.86.094409}{Phys. Rev. B \textbf{86}, 094409, (2012).}
	
	
	
	\bibitem{Landau}
	L.~D. Landau and E.~M. Lifshitz, 
	On the theory of the dispersion of magnetic permeability in ferromagnetic bodies, {Physik. Zeits. Sowjetunion \textbf{8}, 153 (1935)}.
	
	\bibitem{Hubert}
	Alex Hubert, Rudolf Schäfer, Magnetic Domains, \href{https://doi.org/10.1007/978-3-540-85054-0}{Springer Berlin, Heidelberg, 978-3-540-64108-7 (1998).}
	
	
	
	
	\bibitem{Magnoom}
	Kiselev, N. S. \href{https://github.com/n-s-kiselev/magnoom}{Magnoom software (2016).}
	
	\bibitem{Savchenko}
	Savchenko, A. S., Kuchkin, V. M., Rybakov, F. N., Blugel, S., and Kiselev, N. S.  \href{doi:10.1063/5.0097651}{Chiral standing
		spin waves in skyrmion lattice. APL Materials \textbf{10}, 071111 (2022).}
	
	\bibitem{Supplement}
	See the Supplemental Material at [URL] for the weak-coupling limit in the case $N=4$; the wave function ansatz for the ground state in the case $N=4$; the derivation of the intershell interaction Hamiltonian; the asymptotic behavior of the quantum BP profile, its ansatz and details of the numerical simulations in the $\mathbb{S}_{3}$-micromagnetic model; the derivation of the SANS cross-section for the classical and quantum BPs. The Supplemental Material also contains Refs.~\cite{tower, Hesse,Rybakov_thesis,Michels,Svergun,Adams}.
	
	
	
	\bibitem{White}
	S. R. White, Density Matrix Formulation for Quantum Renormalization Groups, \href{https://doi.org/10.1103/PhysRevLett.69.2863}{Phys. Rev. Lett. \textbf{69}, 2863 (1992).}
	
	\bibitem{Haller}
	Andreas Haller, Solofo Groenendijk, Alireza Habibi, Andreas Michels, and Thomas L. Schmidt
	\href{https://doi.org/10.1103/PhysRevResearch.4.043113}{Phys. Rev. Res. \textbf{4} 043113 (2022).}
	
	\bibitem{Dzyaloshinskii}
	I. Dzyaloshinsky,
	A thermodynamic theory of ``weak'' ferromagnetism of antiferromagnetics,
	\href{http://doi.org/10.1016/0022-3697(58)90076-3} {J. Phys. Chem. Solids \textbf{4}, 241 (1958)}.
	
	\bibitem{Moriya}
	T. Moriya,
	Anisotropic superexchange interaction and weak ferromagnetism,
	\href{http://doi.org/10.1103/PhysRev.120.91} {Phys. Rev. \textbf{120}, 91 (1960)}.
	
	\bibitem{Skyrm}
	T.H.R. Skyrme, A Non-Linear Field Theory, \href{https://doi.org/10.1098/rspa.1961.0018}{Proc. R. Soc. Lond. A \textbf{260}, 127 (1961).}
	
	
	
	
	\bibitem{tower}
	O.~M.~Sotnikov, E.~A.~Stepanov, M.~I.~Katsnelson, F.~Mila, and V.~V. Mazurenko, Emergence of Classical Magnetic Order from Anderson Towers: Quantum Darwinism in Action,
	\href{https://doi.org/10.1103/PhysRevX.13.041027}{Phys. Rev. X \textbf{13}, 041027 (2023).}
	
	
	\bibitem{Hesse}
	Hesse, K., Sloan, I.H., Womersley, R.S., Numerical Integration on the Sphere. In: Freeden, W., Nashed, M., Sonar, T. (eds) Handbook of Geomathematics. \href{https://doi.org/10.1007/978-3-642-54551-1_40}{Springer, Berlin, Heidelberg, (2015).} 
	
	\bibitem{Rybakov_thesis}
	P. Rybakov, Topological excitations in field theory models of superconductivity and magnetism, Ph.D. thesis, KTH Royal Institute of Technology, Stockholm (2021).
	
	\bibitem{Michels}
	A. Michels, Magnetic Small-Angle
	Neutron Scattering: A Probe for Mesoscale Magnetism Analysis, Oxford University Press, Oxford, (2021).
	
	\bibitem{Svergun}
	D.~I. Svergun and M.~H.~J. Koch, Rep.
	Prog. Phys. \textbf{66}, 1735 (2003).
	
	\bibitem{Adams}
	M. P. Adams, E. P. Sinaga, H. Kachkachi, and A. Michels, Phys. Rev. B \textbf{109}, 024429 (2024).
	
\end{thebibliography}
\end{document}